# Femtosecond imbalanced time-stretch spectroscopy for ultrafast gas detection

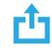

View Online      Export Citation      CrossMark

Zhen Zhang,[1] Haiyun Xia,[1,2,3,a)] (ID) Saifen Yu,[1] Lijie Zhao,[1] Tianwen Wei,[1] and Manyi Li[1]

**AFFILIATIONS**

[1]CAS Key Laboratory of Geospace Environment, University of Science and Technology of China, Hefei 230026, China
[2]Hefei National Laboratory for Physical Sciences at the Microscale, University of Science and Technology of China, Hefei 230026, China
[3]CAS Center for Excellence in Comparative Planetology, University of Science and Technology of China, Hefei 230026, China

[a)]Author to whom correspondence should be addressed: hsia@ustc.edu.cn

**ABSTRACT**

Dual-comb spectroscopy is a promising method for precise optical spectrum analysis with fast data acquisition speed. Here, avoiding using a dual-comb source, femtosecond imbalanced time-stretch spectroscopy with a simple optical layout is proposed and demonstrated. Time-stretch interferometry from one femtosecond laser builds mapping from the optical frequency domain to the radio frequency regime. In experiment, the absorption line of a hydrogen cyanide cell is encoded in the probing arm of a Mach–Zehnder interferometer (MZI). The down-converted radio frequency comb is transformed from a periodically chirped waveform, which is the interferogram of the MZI with different dispersion values on two arms. In a single measurement, the optical frequency comb with a span of 112.5 GHz is down-converted to a range of about 20.8 GHz in the radio frequency domain with a comb spacing equal to the laser repetition frequency of 100 MHz. By turning the optical filter, a spectrum range around 2 THz is analyzed. The acquired optical spectrum resolution is 540 MHz.

*Published under license by AIP Publishing.* https://doi.org/10.1063/1.5143790

Ultrafast spectroscopy is a crucial tool for understanding the substance composition, molecular evolution, and kinetics in not only the fundamental sciences of physics, chemistry, and biomedicine, but also in applied domains of gas tracing and leakage warning. As the widely used spectroscopy in the past few decades, Fourier transform spectroscopy (FTS) provides a broadband spectrum and well-resolved data in nonintrusive diagnostics of molecular structures in various media. Recently, researchers have been focusing on new methods with characteristics of fast acquisition time, high sensitivity, high resolution, and broad spectral bandwidth. With the revolution of optical frequency (OF) metrology brought by stabilized optical frequency combs, dual-comb spectroscopy[1–6] (DCS) emerges as a powerful scheme, covering the above characteristics and equipping with no moving components. DCS with two coherent optical frequency combs holds much promise for laboratorial precision spectroscopy[7,8] and regional comprehensive sensing.[9,10] However, in the process of advancing to practical applications, it is hindered by the complexity of two optical combs. Despite daunting technical challenges, much of the effort of research communities is devoted to enhancing the performance of the dual-comb laser for a compact and stable DCS, incorporating different techniques, such as microresonators,[11–14] electro-optic modulation,[4] and one cavity comb.[15,16]

Another ultrafast technique called time-stretch[17] plays an important role not only in ultrafast single-shot spectroscopy,[18–22] imaging,[23–25] ranging,[26,27] and other measurements of non-repetitive and statistically rare signals[28] but also in ultra-wideband[29] and chirped[30,31] microwave pulse generation for high speed communications and radars. Time-stretch builds a time-to-frequency mapping by introducing group delay dispersion (GVD) using various dispersion devices, such as optical fiber, chirped fiber Bragg gratings, and angular dispersion devices. However, time-stretch often suffers from a low signal-to-noise ratio (SNR) due to the high insertion loss of the adopted dispersion device. So in most cases amplifiers are employed to ensure real-time detection.[17,18] On the other hand, the SNR can be improved by incorporating heterodyne detection,[32–34] where the weak signal is mixed with a strong local oscillator, resulting in a radio frequency (RF) beat signal. For example, in the ultrafast ranging system,[26,27] a non-chirped RF waveform is generated for ultrafast ranging purposes. However, due to its narrow-band RF spectrum, the spectrum information encoded in the envelope is hard to acquire. Interestingly, by introducing imbalanced dispersion in the Mach–Zehnder interferometer (MZI), a chirped interferogram with ultra-wideband (UWB) RF is generated, where the information carried on the optical spectrum is down-converted to the UWB-RF spectrum.

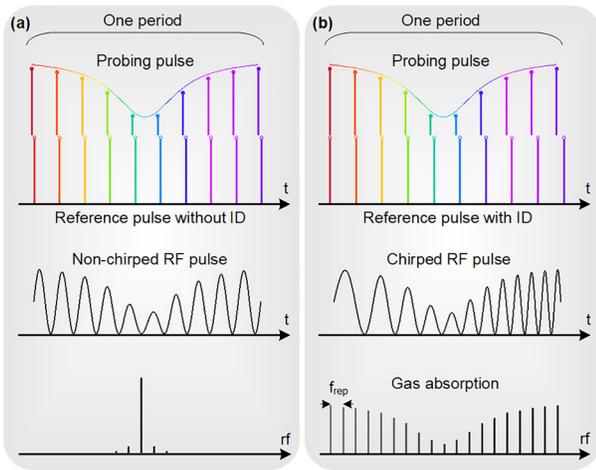

**FIG. 1.** Comparison between time-stretch heterodyne detection with imbalanced dispersion and without imbalanced dispersion (ID). (a) Without imbalanced dispersion, t: time. (b) With imbalanced dispersion. The figures from top to bottom are frequency comb teeth in the stretched probing pulse, frequency comb teeth in the stretched reference pulse, interferograms after heterodyne detection, and RF combs.

As shown in Fig. 1, in the time-stretch heterodyne detection, the probing and reference pulses are split from one femtosecond laser, so that the two pulses have intrinsic coherence. If there is no imbalanced dispersion in the MZI [Fig. 1(a)], a non-chirped RF waveform is generated and the RF spectrum has a narrow bandwidth, while in Fig. 1(b), an imbalanced dispersion is added to the reference pulse. Within one period, the time spacing between neighboring frequency teeth is different in two arms. The heterodyne detection produces a chirped RF pulse, which carries the spectrum information. Thus, the gas absorption feature encoded in the optical spectrum can be retrieved in the envelope of the UWB-RF combs. Here, femtosecond imbalanced time-stretch spectroscopy (FITSS) is implemented by using one femtosecond laser, inheriting features from both time-stretch and DCS, such as downconversion ability from the OF domain to the RF domain, high sensitivity, and ultrafast data-acquisition.

The optical layout of FITSS is shown in the left-dashed line box of Fig. 2. The spectra and pulses at different stages [labeled as (a)–(h)] are shown in the right-dashed line box. A femtosecond laser with stabilized pulses with a repetition rate of 100 MHz serves as the laser source (a). The femtosecond pulse is tailored in the spectrum domain to cover the spectrum range of one absorption line by a programmable filter with a bandwidth of 0.9 nm (b). Then, the pulse is stretched (c) in the time-stretch module consisting of cascaded DCFs with a sufficient dispersion value and EDFAs to compensate the power loss. The optical pulse after dispersion is illustrated in the time-frequency space, as shown in Fig. 2(c). The total dispersion value of the time-stretch module is 14.5 ns/nm. The pulse duration is stretched from tens of femtoseconds to about ten nanoseconds. In order to suppress the sideband effects due to the rectangular window function introduced by the programmable filter, an intensity modulator driven by a waveform generator is employed to tailor the pulse in the time domain with the Gaussian shape (d), as shown in Fig. 2(d). Then, the pulse is split into two pulses in the MZI [Figs. 2(e) and 2(f)]. The reference pulse transmits through an additional DCF with a dispersion of 3.3 ns/nm, resulting in different time-to-frequency mapping functions relative to the probing pulse, as shown in Fig. 2(e). The probing pulse transmits through a $H^{13}CN$ gas cell (48 cm, 100 Torr) with the spectrum shown in Fig. 2(f). The time delay between the two pulses is adjusted by using an optical delayer. The MZI is placed inside a temperature controlled box with temperature fluctuations less than 0.01 centigrade and vibration isolation for stabilizing the phase difference between the two arms. In this work, the system is calibrated with no gas under test in the cell, and then it can be used for gas analysis. In experiment, we find that ambient temperature variance will not only change the optical length difference between the two arms of the interferometer but also affect the coherent detection efficiency by changing the polarization states of devices in the system. Thus, temperature control is used to improve the system stability. After going through the polarization controller and in-line polarizers, the pulses are combined by a coupler for heterodyne detection. A chirped interferogram [Fig. 2(g)] with a period of 10 ns is detected using a photodiode and recorded using an oscilloscope with a sampling rate of 80 Giga samples per second. The bottom interferogram in Fig. 2(g) is formed without the H$^{13}$CN gas cell. Finally, by performing fast Fourier transformation (FFT) on the periodically chirped interferogram, the absorption feature is shown in the RF comb [Fig. 2(h)].

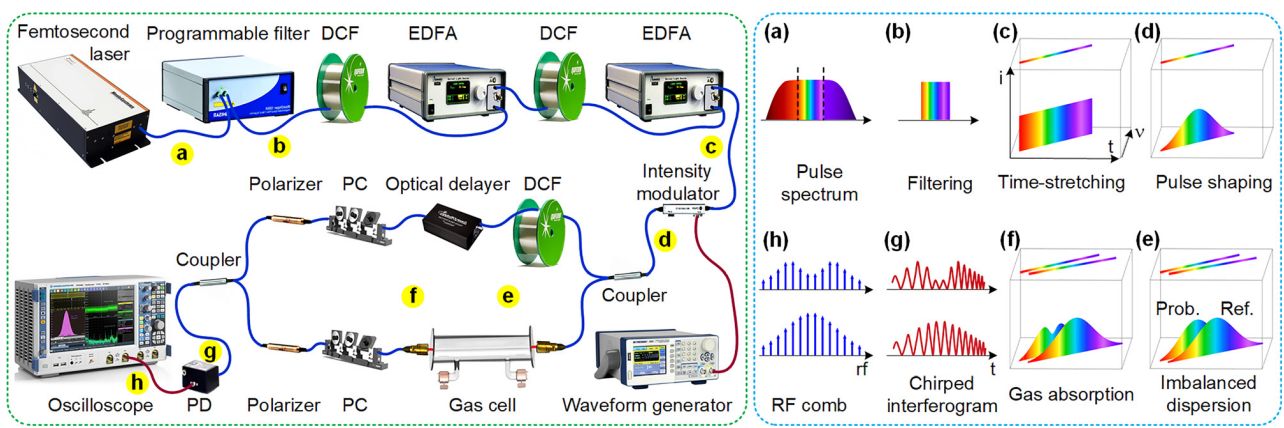

**FIG. 2.** The optical layout of FITSS (in the left-dashed line box) and the spectra and pulses (in the right-dashed line box) at different stages [labeled as (a)–(h)]. (a) The femtosecond laser spectrum is tailored to (b) the range of interest by a programmable filter. (c) The stretched laser pulse shown in the time-frequency space. t: time; $\nu$: optical frequency; And i: intensity. (d) Gaussian shaped pulse. (e) Probing pulse and reference pulse in the MZI without $H^{13}CN$ gas. (f) Pulses after gas absorption in the probing arm. (g) Electric interferogram on the detector. (h) FFT results in the oscilloscope. DCF: dispersion compensation fiber; EDFA: erbium-doped fiber amplifier; PC: polarization controller; and PD: photodiode.

By introducing the time-to-frequency mapping function, the imbalanced dispersion in the FITSS can be described in detail. After passing through a DCF, the femtosecond pulse is dispersed and stretched, and the spectral information maps to the time domain by GVD. FITSS is a time-stretch interferometry, where the dispersion values in the two arms are different. Based on the instantaneous frequency concept in time-stretch,[27] considering dispersion up to the third-order, the instantaneous frequency $\omega$ of the probing pulse and reference pulse can be expressed as

$$\omega_p(t) = t/\beta_2 L_0 - \beta_3 L_0 t^2/2(\beta_2 L_0)^3, \tag{1}$$

$$\omega_r(t) = (t - \tau)/\beta_2(L_0 + L_I) - \beta_3(L_0 + L_I)(t - \tau)^2/2[\beta_2(L_0 + L_I)]^3, \tag{2}$$

where $\omega$ is the angular frequency relative to the center frequency of the pulse, $t$ is the relative group delay time, $L_0$ is the length of DCF in the time-stretch module in Fig. 2, $L_I$ is the length of DCF for imbalanced dispersion in the reference arm, and $\tau$ is the time delay between the two arms. $\beta_2$ and $\beta_3$ are the second-order and third-order mode-propagation constants, respectively.

The beat frequency can be expressed as $\omega_{rf}(t) = |\omega_p(t) - \omega_r(t)|$. The beat frequency can be divided into two parts: the linear chirp part introduced by the imbalanced dispersion and the residual part associated with the third order dispersion, which can be written, respectively, as

$$\omega_I(t) = t/\beta_2 L_0 - (t - \tau)/\beta_2(L_0 + L_I), \tag{3}$$

$$\omega_T(t) = \beta_3(L_0 + L_I)(t - \tau)^2/2[\beta_2(L_0 + L_I)]^3 - \beta_3 L_0 t^2/2(\beta_2 L_0)^3. \tag{4}$$

In this experiment, the system parameters include $\beta_2 = 210$ ps$^2$/km, $\beta_3 = -1.289$ ps$^3$/km, $L = 88$ km, and $L_I = 20$ km. The total dispersion values of the time-stretch module and imbalanced dispersion are about 14.5 ns/nm and 3.3 ns/nm, respectively. The spectrum span of single detection is 0.9 nm, so that the pulse is stretched to about 13 ns in the time-stretch module. Given the time delay of 25 ps, the frequency chirped by the imbalanced dispersion $\omega_I(t)$ ranges from 1.1 GHz to 21.91 GHz, while the frequency chirp due to the third-order dispersion (TOD) is in the range of 3.73 kHz to 82.19 MHz. For simplicity, in the rest of this work, ignoring the effect of TOD only introduces a relative error less than 0.39%.

The DCF lengths $L_0$ and $L_I$ determine the RF spectrum span when the optical spectrum span is constant, and the time delay $\tau$ determines the center frequency. The OF-to-RF compression ratio is defined as the ratio between the optical spectrum range and the radio spectrum range. It is related to the length ($L_0$) of DCF in the time-stretch module and the length ($L_I$) of DCF for imbalanced dispersion in the reference arm. This ratio can be enhanced by using either longer

$L_0$ or shorter $L_I$. Then, the OF working bandwidth in single detection can be expanded, or the RF bandwidth can be reduced in a constant OF spectrum range.

As shown in Fig. 2(f), the probing and reference pulses are separated from a Gaussian shaped pulse, and then the probing pulse is absorbed after passing through the H$^{13}$CN gas cell, whose absorption function can be seen as a Voigt function. So the spectrum intensity in two arms of the MZI can be expressed as $I_p(\omega) = G_G(\omega)[1 - G_V(\omega)]$ and $I_r(\omega) = G_G(\omega)$, where $G_G(\omega)$ and $G_V(\omega)$ represent the Gaussian function and Voigt function, respectively. The complex spectrum can be expressed as $a_p(\omega) = [I_p(\omega)]^{0.5}\varphi(j\omega)$ and $a_r(\omega) = [I_r(\omega)]^{0.5}\varphi(j\omega)$. The chromatic dispersion up to group delay dispersion performs an optical Fourier transform to the pulse.[35] The reference pulse transmits through an optical delayer and an additional DCF. So considering the dispersion up to group delay dispersion, the complex pulse[35] can be expressed as

$$a_p(t) = h_p \exp\left(jt^2/2\beta_2 L_0\right)[a_p(\omega)]_{\omega = t/\beta_2 L_0}, \tag{5}$$

$$a_r(t) = h_r \exp\left[j(t - \tau)^2/2\beta_2(L_0 + L_I)\right][a_r(\omega)]_{\omega = (t - \tau)/\beta_2(L + L_I)}, \tag{6}$$

where $h_p = H_0(j2\pi\beta_2 L_0)^{-0.5}\exp(-j\beta_0 L_0)$ and $h_r = H_0[j2\pi\beta_2(L_0 + L_I)]^{-0.5}\exp[-j\beta_0(L_0 + L_I)]$ are the complex amplitude of the two complex pulse. $H_0$ is a constant of the fiber transmission rate, and $\beta_0$ is the zero-order mode-propagation constant. For simplicity, the following equations are introduced $C_p = h_p \exp(jt^2/2\beta_2 L_0)$ and $C_r = h_r \exp[j(t - \tau)^2/2\beta_2(L_0 + L_I)]$. Hence, Eqs. (5) and (6) can be expressed as

$$a_p(t) = C_p[G_G(t/\beta_2 L_0)]^{0.5}[1 - G_V(t/\beta_2 L_0)]^{0.5} \exp\left[j\varphi(t/\beta_2 L_0)\right], \tag{7}$$

$$a_r(t) = C_r\{G_G\left[(t - \tau)/\beta_2(L_0 + L_I)\right]\}^{0.5} \times \exp\{j\varphi[(t - \tau)/\beta_2(L_0 + L_I)]\}. \tag{8}$$

The probing and reference pulses are combined by a coupler and fed to a PD. The interference term of the detector response photocurrent can be derived from Eqs. (7) and (8),

$$i(t) = [G_E(t)]^{0.5}[1 - G_V(t/\beta_2 L_0)]^{0.5} \cos\left[\omega_{rf}(t)t\right], \tag{9}$$

where $G_E(t) = |C_p||C_r|G_G[(t - \tau)/\beta_2(L_0 + L_I)]G_G(t/\beta_2 L_0)$ is the envelope without the absorption feature. $|C_p| = H_0(2\pi\beta_2 L_0)^{-0.5}$ and $|C_r| = H_0[2\pi\beta_2(L_0 + L_I)]^{-0.5}$ are constants. $\omega_{rf}(t) = |\omega_p(t) - \omega_r(t)|$ is the beat frequency. Using Eqs. (1) and (2), the beat frequency can be expressed as $\omega_{rf}(t) = t/\beta_2 L_0 - (t - \tau)/\beta_2(L_0 + L_I)$. If there is no imbalanced dispersion, i.e., $L_I = 0$, the beat frequency will be a constant $\omega_{rf} = \tau/\beta_2 L_0$. While the imbalanced dispersion is employed in FITSS, the beat frequency is chirped from $\omega_{rf}(t_{min})$ to $\omega_{rf}(t_{max})$. By performing FFT to $i(t)$ in Eq. (9), the absorption feature is retrieved in the envelope of the wideband RF spectrum.

The former DCS establishes a direct link between the OF domain and the RF domain by employing two coherent combs with slightly different repetition rates. Here, the spectrum downconversion is realized by using one femtosecond laser and imbalanced dispersion in time-stretch interferometry, making the system simple and thus robust.

In experiment, the periodically chirped interferogram is shown in Fig. 3(a). One should know that the overlap of the stretched pulses in the time domain has no effect on the transformed RF combs. The enlarged figures in Fig. 3(b) show the chirped frequency of the interferograms over one period. The upper interferogram in Fig. 3(b) is acquired when the gas cell is filled with nitrogen for calibration, while the bottom interferogram is acquired when the gas cell is filled with gas under test. The RF combs are obtained by performing FFT on the time-domain waveforms. In order to get a SNR better than 25 dB in the center lines, the interferograms over 25 000 periods are recorded for FFT. However, the minimum acquisition time can be shortened to 10 ns in single-shot detection when the SNR of the system is enhanced. In that case, the RF spectrum is not a comb structure but an envelope of the absorption spectrum. As shown in Figs. 3(c) and 3(d), the reference comb and the signal comb in the RF domain have the same comb spacing, which is equal to the laser pulse repetition frequency. Here, the optical spectrum with a span of 112.5 GHz in the OF domain is down-converted to a range of about 20.8 GHz in the RF domain, which means that the OF-to-RF compression ratio is 5.4.

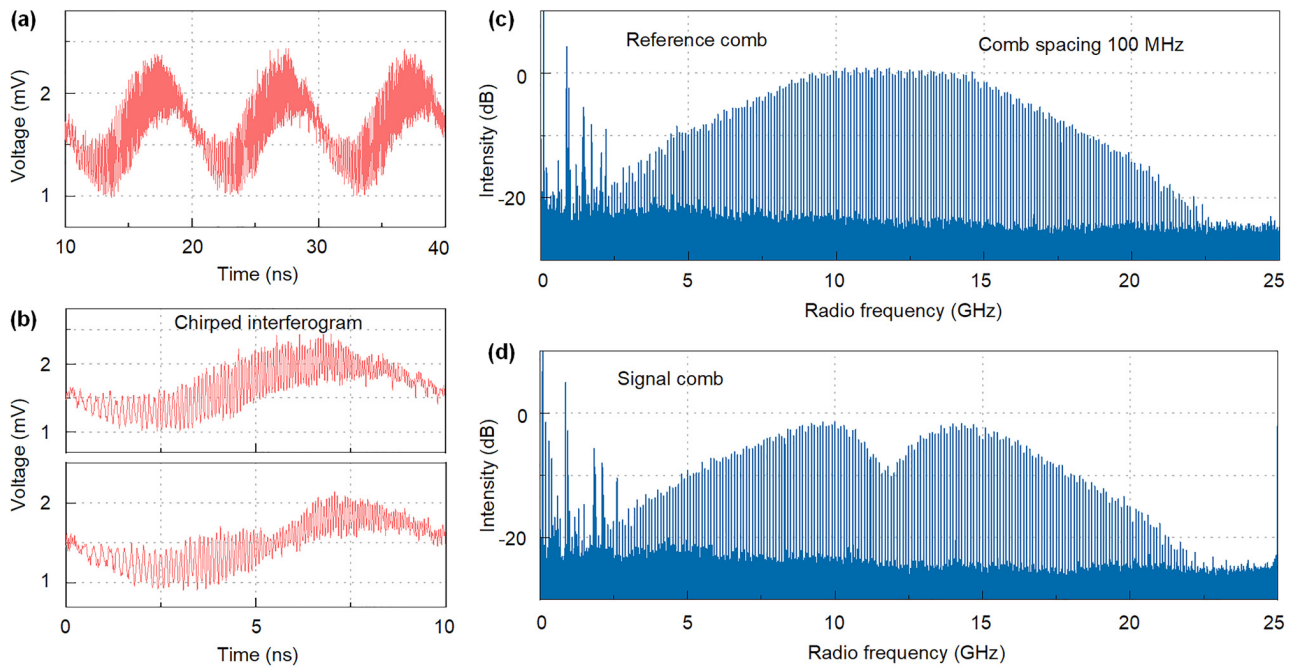

**FIG. 3.** Measured electrical interferograms and transformed RF combs. (a) Periodically chirped interferogram with a period of 10 ns. (b) Enlarged interferograms. Upper: without the absorption feature. Bottom: with the absorption feature of H$^{13}$CN. (c) and (d) RF combs obtained by performing Fourier transform on the temporal interferograms with a recording time of 25 000 periods. The comb without the gas feature functions as a reference.

Finally, after transforming the recorded waveforms, the absorption spectrum is normalized from the peaks of the signal and reference combs. In Fig. 4(a), the measured FITSS spectrum (black circles) in the $2\nu_3$ band of H$^{13}$CN from 192.36 THz to 194.34 THz is compared with the absorption spectrum directly measured using an optical spectrum analyzer (OSA). A wide spectrum is acquired by tuning the programmable filter. 20 absorption features are assembled. The line-by-line overlay of the measured data from FITSS and OSA shows good uniformity in the overall profiles. Some negative numbers appear in the normalized absorption feature due to the low SNR sample points on the edges of the RF comb. The residual difference between the two absorption spectra is shown as blue dots in Fig. 4(a). Then, the raw data of FITSS are fitted with a Voigt function [Fig. 4(b)]. The final FITSS result refers to the fitted curve that is compared with the OSA spectrum in Fig. 4(c), which confirms the wavelength precision and absorption intensity accuracy of this method. The acquisition time of one absorption line over 0.9 nm is about 1s by using OSA, while it is 250 $\mu$s by averaging 25 000 periods in FITSS. Most peaks of the absorption lines in OSA are slightly lower than those in FITSS due to the limited resolution (2.5 GHz) of the OSA. The spectrum resolution of FITSS is directly determined by the pulse repetition rate. In this experiment, a pulse repetition frequency of 100 MHz corresponds to an optical frequency resolution of 540 MHz at 1550 nm. The resolution is the product of the pulse repetition rate and the compression factor of 5.4. This implies that one can improve the resolution at the sacrifice of detection speed.

In conclusion, instead of using two stabilized comb sources, FITSS is demonstrated to down-convert the molecular spectrum from the OF domain to the RF domain by utilizing one femtosecond laser, incorporating the time-stretch interferometer with imbalanced dispersion. Passive optical components for the telecommunication industry are predominantly used. The gas absorption feature is converted to the RF domain that has a comb structure. FITSS obtained the gas absorption feature from the peaks of the RF comb instead of the time domain. Since the OF spectrum range is compressed to the RF spectrum range by a compression factor of 5.4, the spectrum resolution of 540 MHz is the product of the RF comb spacing (the pulse repetition rate is 100 MHz) and the compression factor. In addition, benefiting from the time-stretch technique, FITSS is characterized by an all-fiber compact structure, high detection sensitivity and resolution, and ultra-fast data acquisition speed. With further system development, FITSS can be expanded to practical applications of non-intrusive and non-repetitive gas monitoring in chemical reaction and remote sensing of gas concentrations in real-time. In this work, the filtered spectrum range of 0.9 nm is used to detect one absorption line of the gas under test. For the purpose of different gas species detection in the C and L optical communication bands, such as H$_2$S, CO$_2$, CH$_4$, and H$_2$O, one can tailor the femtosecond laser spectrum to the range of interest. Furthermore, simulated Raman amplification[17,23] is superior to compensate the signal power loss in the photonics time-stretch system, allowing larger dispersion in the system. Thus, the OF working bandwidth can be expanded. With an enhanced SNR, even single-shot detection can be achieved.

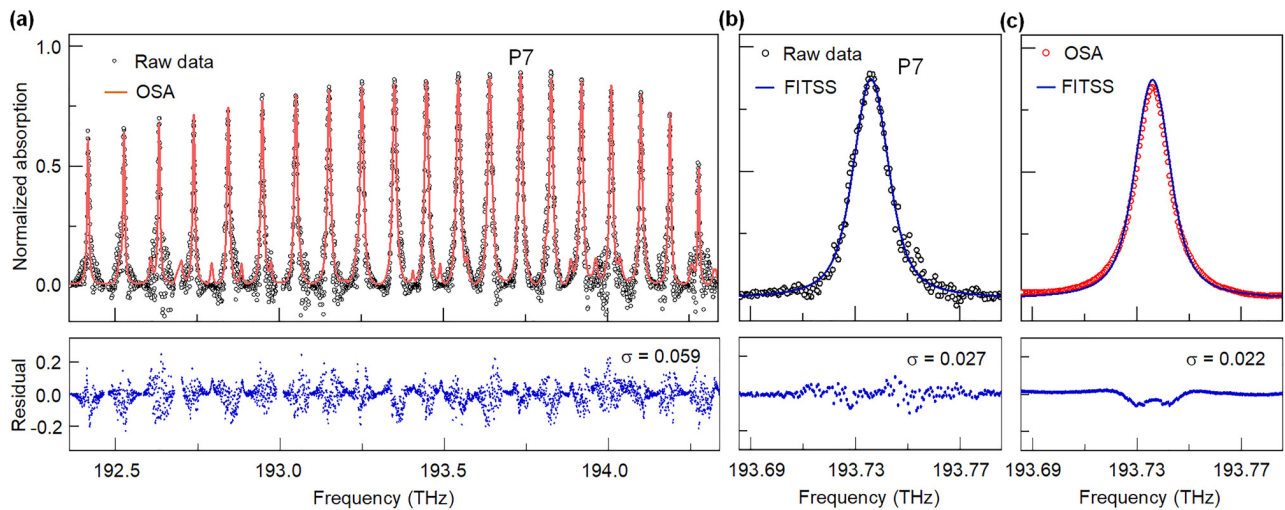

FIG. 4. Measured molecular absorption spectra. (a) Absorption features in the $2\nu_3$ band of H$^{13}$CN are assembled by 20 absorption lines. The overlay of the raw data and the spectrum directly measured using optical spectrum analyzer (OSA) show line-by-line matching. The standard deviation of the residuals is 0.059. (b) The result of FITSS is obtained by fitting the raw data with a Voigt function. The resolution of FITSS with a laser repetition rate of 100 MHz is 545 MHz. (c) Comparison of the results from FITSS and OSA.

This work was supported by the Anhui Initiative in Quantum Information Technologies.